\documentclass[10pt]{article}
\usepackage[a4paper, total={6.5in, 8in}]{geometry}

\usepackage[T1]{fontenc}
\usepackage{lmodern}
\usepackage[utf8]{inputenc}
\usepackage{fancyhdr}
\usepackage{graphicx}
\usepackage{pifont}
\usepackage{dcolumn}
\usepackage{bm}
\usepackage{caption,subcaption}

\usepackage{amsthm}
\usepackage{latexsym}
\usepackage{graphics}
\usepackage{amssymb}
\usepackage{amsfonts}
\usepackage{mathrsfs}
\usepackage{bm}
\usepackage{amsmath}
\usepackage{color}
\usepackage[dvipsnames]{xcolor}
\usepackage{arydshln} 
\usepackage[title]{appendix}

\usepackage{authblk}

\usepackage{cancel}

\RequirePackage{graphicx}
\RequirePackage{flushend}
\RequirePackage[colorlinks,citecolor=blue,urlcolor=blue,linkcolor=blue]{hyperref}

\usepackage{mathtools}  

\newtheorem{definition}{Definition}[section]

\begin{document}


\title{Vacuum energy of  scalar fields on spherical shells with general matching conditions}
\bigskip

\author[1]{Guglielmo Fucci\footnote{e-mail: fuccig@ecu.edu}}
\author[2]{C\'esar Romaniega\footnote{e-mail: csromaniega@gmail.com}}

\affil[1]{\small{Department of Mathematics, East Carolina University}\\ \small{Greenville, North Carolina 27858, USA.}}
\affil[2]{\small{Departamento de F\'isica Te\'orica, At\'omica y \'Optica, Universidad de Valladolid}\\\small{Valladolid, 47011, Spain.}}
\bigskip

\bigskip


\maketitle

\begin{abstract}
In this work we analyze the spectral zeta function for massless scalar fields propagating in a $D$-dimensional flat space under
the influence of a shell potential. The shell potential is defined in terms of the two-interval self-adjoint extensions of the Hamiltonian 
describing the dynamics of the scalar field. After performing the necessary analytic continuation, we utilize the spectral zeta function 
of the system to compute the vacuum energy of the field.

\end{abstract}
\section{Introduction}\label{sec:I}
The Casimir effect is one of the most important predictions of quantum field theory. The effect was first theorized by Casimir 
in his famous 1948 paper \cite{casimir1948attraction}. However, due to the small size of the forces involved, it became evident that a direct experimental verification of the effect would be quite challenging.
Fortunately, recent progress in experimental designs and measuring instruments confirms a close agreement between theoretical predictions of the Casimir effect and the experiments conducted \cite{bordag2009advances,zhao2019stable,munkhbat2021tunable}. 
Apart from its theoretical interest, the Casimir effect, which in essence is a macroscopic manifestation of the vacuum energy of quantum fields, has been investigated as a mechanism for designing contact-free nanomachines. The main goal of these applications is to devise microscopic configurations in which the Casimir-Lifshitz forces prevent stiction between two separate sections of the configuration and provide a stable equilibrium \cite{zhao2019stable}, Other applications include the creation of self-assembling optical microcavities and polaritons \cite{munkhbat2021tunable}.
In all these experiments, one measures explicitly the interaction energy between separates bodies. These measurements are possible because the  Casimir energy and the corresponding force are finite and well-defined \cite{bordag2009advances}. To better understand this point, it is important to mention that the formal expression for the Casimir energy leads, naturally, to a divergent quantity. Meaningful results can be obtained only 
after a suitable renormalization of the energy \cite{bordag2009advances}, a procedure that is often used in other areas of quantum field theory.
The divergences that arise while computing the Casimir energy are proportional to certain coefficients of the asymptotic expansion of the heat kernel associated with the Laplace-type operator describing the dynamics of the quantum field \cite{vassilevich2003heat,kirsten2001spectral}.
When considering the Casimir energy between two separate bodies, these heat kernel coefficients are independent of the distance between the bodies. This implies that while the energy between them may or may not be finite, the force between them is, in fact, a finite quantity \cite{bordag2009advances,kirsten2001spectral}.  

The Casimir energy of single bodies, referred to as Casimir self-energy, is, on the other hand, a more involved concept since surface divergences usually appear \cite{cavero2006local}, which lead to no natural way of renormalizing quantities such as the self-stress \cite{bordag2009advances}.
To illustrate this issue, one can consider a sphere contained within another object ,such as a cavity, for instance. 
The total vacuum energy $E_0$ of this configuration can be written as a sum, after subtracting the contribution from the free space,
\begin{equation}\label{00}
	E_0=E_1+E_2+E_\text{int},
\end{equation}
where $E_1$ and $E_2$ represent the self-energies of the first and second body, respectively and $E_\text{int}$ denotes the interaction energy.
Now, the self-energies $E_1$ and $E_2$ are comprised of both a finite and a divergent part. These depend on the geometry of the objects, such as the radius in the specific case of the sphere. The interaction energy $E_\text{int}$ contains, instead, information about the {\it relative} geometry of the two objects such as the distance from each other. Since the Casimir pressure on the sphere is defined as the variation of the energy with respect to the radius, and since the divergent part of the self-energy of the sphere generally depends on its radius, the pressure will, consequently, be divergent. This divergence can be eliminated, in the case of a massive field, by using the well-known large mass renormalization \cite{bordag2009advances}. It is important to mention that accurate measurements of the Casimir pressure are much more challenging and, to the authors' knowledge, there has not been any experimental attempts yet.   
If the two objects lie outside of each other, the Casimir energy has the same form as in (\ref{00}) but the force between them is defined as the variation of the total energy with respect to the distance between the objects (although the force can also be defined in this way when one object is inside the other, see for instance \cite{rahi2009scattering}). Since, in most cases, the divergent parts of the energies $E_1$, $E_2$, and $E_\text{int}$ is independent of this distance, the Casimir force is well-defined (even if the energy might be divergent \cite{bordag2009advances,kirsten2001spectral,fucci2021piston}).

In this paper, we focus on the analysis of the Casimir pressure on a single sphere, paying particular attention to those cases where this pressure can be defined unambiguously, thus continuing the work of \cite{cavero2021casimir}, which focused on the an analysis of the interaction energy for a massless scalar field in the presence of two concentric $\delta$-$\delta'$ spheres. 
It is important to point out that there are only a few configurations for which the Casimir self-energy is well-defined without the need for renormalization \cite{milton2008local}. Some important examples of such cases include the dilute limit for spheres and cylinders \cite{bordag1999ground,cavero2005casimir}, a magnetodielectric object where the speed of light is the same inside and outside \cite{klich1999casimir,milton1999mode}, a perfectly conducting spherical or cylindrical shell \cite{boyer1968quantum,deraad1981casimir}, 
and the $\delta$-potential weak limit for massless scalar fields \cite{milton2004casimirdelta}. 

The configuration we consider here consists of a single sphere described by a shell potential which generalizes the $\delta$ and $\delta$-$\delta'$ results obtained in previous papers. The shell potential is defined in terms of self-adjoint extensions of the Laplacian on two intervals and will be described in detail in the next section. Our proposal of describing the sphere by means of a general shell potential has some clear advantages compared to the previous approaches. First, as we will see in the next section, the general shell potential contains a number of free parameters that, once fixed, describe a particular potential. Because of this freedom, we can explicitly compute the divergent part of the Casimir energy and 
analyze the class of potentials for which this divergence vanishes. This will provide valuable information regarding the types of boundary conditions that one can impose on the sphere which lead to a finite energy. Second, this potential is simple enough that the analysis of the spectral zeta function and the Casimir energy can be carried out explicitly but complicated enough to provide results that are actually of interest.       

In order to analyze the Casimir energy of our system, we will employ the spectral zeta function regularization method \cite{kirsten2001spectral}. 
We represent the spectral zeta function of the operator describing the dynamics of the quantum field in terms of a contour integral
and we analytically continue this expression, by utilizing the methods described in \cite{kirsten2001spectral}, to a region of the complex plane that is appropriate for the computation of the Casimir energy. 

The outline of the paper is as follows. In the next section, we use the theory of self-adjoint extensions of a Sturm-Liouville operator on two 
intervals in order to define a general $\delta$-shell potential. In Section \ref{SectZet}, we perform the analytic continuation of the spectral zeta function and in Section \ref{casEner} we compute the Casimir energy and we provide explicit numerical results for some configurations 
for which the energy is a well-defined quantity. The Conclusions summarize the main results and point to future directions of research.

\section{The shell potential}

In this work we consider a massless scalar field propagating in a $D$-dimensional flat space under the influence 
of a spherically symmetric potential $V(|{\bf x}|)$, ${\bf x}\in\mathbb{R}^{D}$, with $\textrm{Supp}(V)=S_{R}$ where $S_{R}$ denotes the $D$-dimensional sphere of radius $R$.  
We will henceforth refer to $V(|{\bf x}|)$ as a {\it shell} potential due to the fact that its support is $S_{R}$. The precise definition of the shell potential 
is given in terms of the rotationally invariant self-adjoint extensions in $L^{2}(\mathbb{R}^{D},d{\bf x})$ of the following
minimal operator 
\begin{equation}\label{oper1}
H_V=-\Delta+V(|{\bf x}|)\;,
\end{equation}
where $\Delta$ represents the $D$-dimensional Laplacian, with domain 
\begin{equation}
D(H_V)=\{\varphi\in L^{2}({\cal R}^{D},d{\bf x})|\varphi,\varphi'\in AC_{\textrm{loc}}({\cal R}^{D}), \textrm{supp}\,\varphi\subset{\cal R}^{D}, H_V\varphi\in L^{2}({\cal R}^{D},d{\bf x})\}\;,
\end{equation}
where ${\cal R}^{D}=\mathbb{R}^{D}\backslash S_{R}\cup\{0\}$.
In order to characterize the rotationally invariant self-adjoint extensions of $H_V$ we decompose $L^{2}(\mathbb{R}^{D},d{\bf x})$ in terms of the angular momenta as follows \cite{antoine87,albeverio,gallone}
\begin{equation}
L^{2}(\mathbb{R}^{D},d{\bf x})=\bigoplus_{l=0}^{\infty}L^{2}((0,\infty),r^{D-1}dr)\otimes\textrm{span}\{Y_{l}^{-l},\ldots,Y_{l}^{l}\}\;,
\end{equation}
where $Y_{l}^{m}$, with $-l\leq m\leq l$, are the hyperspherical harmonics in $L^{2}(\mathcal{S}^{D-1},d\Omega)$ with $\mathcal{S}^{D-1}$ being the $(D-1)$-dimensional unit sphere.
With respect to this decomposition, the operator $H_V$ is reduced to 
\begin{equation}\label{operdeco}
H_V=\bigoplus_{l=0}^{\infty}h_{l} \otimes \mathbb{I}\;,
\end{equation} 
where the (minimal) \emph{radial} operator $h_{l}$ in $L^{2}((0,\infty),r^{D-1}dr)$ is defined as
\begin{equation}\label{radoper}
h_{l}=-\frac{1}{r^{D-1}}\frac{d}{dr}\left(r^{D-1}\frac{d}{dr}\right)+\frac{l(l+D-2)}{r^2}\;,
\end{equation}
with domain
\begin{equation}
D(h_{l})=\{\phi\in L^{2}({\cal R},r^{D-1}dr)|\phi,\phi'\in AC_{\textrm{loc}}({\cal R}), \textrm{supp}\,\phi\subset{\cal R}, h_{l}\phi\in L^{2}({\cal R},r^{D-1}dr)\}\;,
\end{equation}
where ${\cal R}=\mathbb{R}_{>0}\backslash \{R\}$. 
We are now in a position to give the definition of a shell potential as follows:
\begin{definition}\label{def}
The shell potential is defined as the interaction which constraints the dynamics of the scalar field $\varphi$ to be governed by a 
given rotationally invariant self-adjoint extension of the operator (\ref{operdeco}) on the two regions $0<|{\bf x}|<R$ and $|{\bf x}|>R$.
\end{definition}

The decomposition performed above implies that the rotationally invariant self-adjoint extensions of (\ref{oper1}) reduce to those of the 
radial operator (\ref{radoper}) on the two intervals $I_{<}=(0,R)$ and $I_{>}=(R,\infty)$ (see e.g. \cite{gallone}). The two-interval 
self-adjoint extensions of a Sturm-Liouville operator (of which (\ref{radoper}) is a particular case) have been treated extensively in 
Chapter 13 of the monograph \cite{zettl}. 
According to \cite{zettl}, all two-interval self-adjoint extensions of (\ref{radoper}) can be characterized in terms of boundary conditions 
imposed on functions $\phi$ in the maximal domain $D_{\textrm{max}}(h_{V,l})=D_{<}(h_{V,l})+D_{>}(h_{V,l})$ of (\ref{radoper}) where 
\begin{equation}
D_{j}(h_{l})=\{\phi\in L^{2}(I_{j},r^{D-1}dr)|\phi,\phi'\in AC_{\textrm{loc}}(I_{j}), h_{V,l}\phi\in L^{2}(I_{j},r^{D-1}dr)\}\;.
\end{equation}
In Appendix \ref{app} we have outlined all two-interval self-adjoint extensions of (\ref{radoper}), and we have indicated that they depend 
on the limit point or limit circle classification of $r=0$. We can, therefore, describe the shell potentials associated to the self-adjoint extensions of
(\ref{radoper}), and consequently of (\ref{oper1}), by analyzing the boundary conditions that characterize the latter.  

According to the definition (\ref{def}) and the results in Appendix \ref{app}, for $D\geq 4$ and $l\geq 0$ and for $D=2,3$ and $l\geq 1$, we have essentially two types of shell potentials: The first type, which is associated with separated boundary conditions at $r=R$ (\ref{bcsep}), consists of a completely \emph{opaque} potential that renders 
the interior and the exterior of $S_{R}$ independent of each other. In other words, scalar fields propagating inside $S_{R}$ cannot 
cross the shell potential (the same behavior occurs for scalar fields on the outside of $S_{R}$).
The second type of potential that can occur, which is associated with coupled boundary conditions at $r=R$ (\ref{bccoup}), involves a \emph{semi-transparent} one, where
fields propagating in the interior of $S_{R}$ can interact with those propagating on the exterior of $S_{R}$.

When $D=2,3$ and $l=0$ the boundary conditions allowed, and hence the shell potentials they define, are more involved as it can be recognized from (\ref{bcthree}). 
However, not all the possible shell potentials can be considered. In fact, in order for the scalar field to interact with the same type of potential for all values of the angular momentum $l$, we have to restrict the possible shell potentials, for $l=0$ in $D=2,3$, only to those that are opaque and semi-transparent.
It is not difficult to prove that the opaque and semi-transparent shell potentials described above form a subset of those generated by the boundary conditions (\ref{bcthree}). 
In fact, since the $\textrm{rank}({\cal A}|{\cal B}|{\cal C})=3$ we can choose the matrices ${\cal A}$, ${\cal B}$ and ${\cal C}$ to have the form
\begin{equation}\label{matrix1}
{\cal A}=\begin{pmatrix}
1 & 0 \\
0 & 0 \\
0 & 0
\end{pmatrix}\;,\quad 
{\cal B}=\begin{pmatrix}
0 & 0 \\
b_{21} & b_{22} \\
b_{31} & b_{32}
\end{pmatrix}\;,\quad
{\cal C}=\begin{pmatrix}
0 & 0 \\
c_{21} & c_{22} \\
c_{31} & c_{32}
\end{pmatrix}\;,
\end{equation}
where the entries $b_{ij}$ and $c_{ij}$ with $i=\{2,3\}$ and $j=\{1,2\}$ are not all zero.
With the choice (\ref{matrix1}), the generalized boundary conditions at the origin become completely decoupled from the 
boundary conditions that can be imposed at $r=R$. The matrix ${\cal A}$ has been chosen so that the $l=0$ component of the scalar field, in two and three dimensions, satisfies the generalized Dirichlet boundary conditions at $r=0$. This condition ensures that the only allowed eigenfunctions
of $h_0$ in $I_{<}$ are Bessel functions of the first kind, matching, in this way, the interior eigenfunctions allowed for higher angular momenta.
In other words, the generalized Dirichlet condition for the zero angular momentum at $r=0$ mimics the limit point condition that
occurs at the origin when considering $l\geq 1$.     
To analyze the boundary conditions at the point $r=R$, we turn our attention to the relations (\ref{cond1}). We notice that the first and third 
equation in (\ref{cond1}) are automatically satisfied, while the second expression leads to the equality
\begin{equation}\label{cond}
\textrm{det}{\cal B}_{23}=\textrm{det}{\cal C}_{23}\;.
\end{equation} 
It is this condition that determines the allowed boundary conditions at the point $r=R$. In fact, if $\textrm{det}{\cal B}_{23}=\textrm{det}{\cal C}_{23}=0$, the requirement that $\textrm{rank}({\cal A}|{\cal B}|{\cal C})=3$ implies that 
the only allowed boundary conditions at $r=R$ are separated (like those in (\ref{bcsep})). If, on the other hand, $\textrm{det}{\cal B}_{23}=\textrm{det}{\cal C}_{23}\neq 0$, then the constraint $\textrm{rank}({\cal A}|{\cal B}|{\cal C})=3$ is automatically satisfied, the submatrix 
${\cal C}_{23}$ is invertible, and the boundary conditions allowed at $r=R$ are the coupled ones (i.e. of the form (\ref{bccoup})). 
We can hence conclude that for $l=0$ and $D=2,3$ the matrices in (\ref{matrix1}) together with the condition (\ref{cond}) represent either the opaque or the semi-transparent shell potentials allowed when $l\geq 1$.

In this work we will focus on the semi-transparent shell potentials because they allow for the interaction of the fields inside and 
outside $S_{R}$. This interaction leads to much more interesting results for the Casimir energy of the system. Opaque shell potentials, on the other hand, are somewhat less interesting since the dynamics of the field in the interior of $S_{R}$ is independent of the one in the exterior of $S_{R}$. This implies, in particular, that the vacuum energy of the scalar field is simply the sum of the vacuum energies one would obtain by considering the interior and exterior regions separately.

\section{The spectral zeta function}
\label{SectZet}

In order to study the vacuum energy 
of the scalar field propagating in $\mathbb{R}^{D}$ under the influence of a shell potential we utilize the spectral zeta function of 
the operator $H_{V}$. By definition \cite{kirsten2001spectral}, the spectral zeta function of $H_{V}$ is defined as
\begin{equation}\label{zeta}
\zeta(s)=\sum_{\lambda_{n}\in\sigma(H_V)}\omega_{n}^{-s}\;,
\end{equation}  
where $\sigma(H_V)$ denotes the spectrum of $H_V$ endowed with the boundary conditions which define the semi-transparent shell potential.
According to the general theory of the spectral zeta function (see e.g. \cite{kirsten2001spectral}), (\ref{zeta}) is well-defined 
for $\Re(s)>D/2$ and can be analytically continued to a meromorphic function with isolated simple poles to the right of the abscissa of
convergence $\Re(s)=D/2$. Once the analytically continued expression for (\ref{zeta}) is obtained, one can use it to analyze, in particular,
the vacuum energy of the system \cite{bordag2009advances,kirsten2001spectral}. 
Due to the spherical symmetry of the system, the spectral zeta function (\ref{zeta}) can be written as
\begin{equation}\label{zeta1}
\zeta(s)=\sum_{l\geq 0}d(l)\lambda_{l}^{-2s}\;,
\end{equation}   
where $\lambda_{l}$ are the eigenvalues of the radial operator (\ref{radoper}) endowed with the coupled boundary conditions (\ref{bccoup}), which
can be written more explicitly as
\begin{eqnarray}\label{bccoupexp}
\left(\begin{array}{c}
	 \phi(R_{+}) \\
	\phi'(R_{+})\\
	\end{array}\right)=e^{-i \mu}\begin{pmatrix}
a & b \\
c & d 
\end{pmatrix} 
\left(\begin{array}{c}
	 \phi(R_{-}) \\
	\phi'(R_{-})\\
	\end{array}\right)\;,\quad ad-bc=1\;,
\end{eqnarray} 
and $d(l)$ represent the degeneracy of each eigenvalue, for $D\geq 2$, 
\begin{equation}\label{degeneracy}
d(l)=\frac{(D+l-3)!}{(D-2)!l!}(D-2+2l)\;.
\end{equation}  
The eigenvalues $\lambda_{l}$ needed to construct the spectral zeta function can be obtained implicitly by imposing the boundary conditions 
(\ref{bccoupexp}) to the interior and exterior solutions of the eigenvalue equation associated with the 
radial operator (\ref{radoper}) (or equivalently of (\ref{app1a})). We would like to point out that the coupled boundary conditions 
(\ref{bccoupexp}) contain, as a particular case, those describing point supported potentials previously considered in the literature.
In fact, by setting $a=d=1$ and $b=0$, the conditions (\ref{bccoupexp}) represent a well-known $\delta$-potential at $r=R$. 
By setting, instead, $d=a^{-1}$ and $b=0$, we obtain the matching conditions describing the $\delta$-$\delta'$ potential at $r=R$ 
which was introduced in \cite{kurasov1996distribution} and studied in different contexts over the years in \cite{gadella2009bound1,munoz2015delta,albeverio2000singular,martin2022solvable}.

In what follows, we will provide the analysis of the spectral zeta function
and vacuum energy for $D\geq 3$.  
By setting $\nu=l+(D-2)/2$, one can write the solution of (\ref{app1a}) on the interior region $I_<$ as
\begin{equation}\label{intsol}
\phi_{\textrm{int}}(\lambda r)=r^{-\frac{D-2}{2}}J_{\nu}\left(\lambda r\right)\;.
\end{equation}
This is the solution allowed at the limit point $r=0$ when $D\geq 4$ and $D=3$ with $l\geq 1$. In addition, (\ref{intsol}) also coincides with the solution that satisfies the generalized Dirichlet boundary condition at limit circle point $r=0$ when $D=3$ and $l=0$. 
The solution in the exterior region $I_>$ can be written as a linear combination of Hankel functions as follows 
\begin{equation}\label{extsol}
\phi_{\textrm{ext}}(\lambda r)=r^{-\frac{D-2}{2}}\left[A_{1}H^{(2)}_{\nu}(\lambda r)+A_{2}H^{(1)}_{\nu}(\lambda r)\right]\;.
\end{equation}
 
By imposing the coupled boundary conditions (\ref{bccoupexp}), with $\mu=0$ for simplicity, on the interior 
and exterior solutions (\ref{intsol}) and (\ref{extsol}) we obtain the following expressions for the coefficients $A_{j}$, with $j=\{1,2\}$ of (\ref{extsol})
\begin{eqnarray}\label{sol}
A_{j}(\lambda R)&=&\frac{(-1)^{j}i \pi\lambda R}{4}\Bigg\{\left[a\left(\lambda H^{(j) \prime}_{\nu}(\lambda R)-\frac{(D-2)}{2R}H^{(j)}_{\nu}(\lambda R) \right)-cH^{(j)}_{\nu}(\lambda R)\right]\phi_{\textrm{int}}(\lambda R)\nonumber\\
&+&\left[b\left(\lambda H^{(j) \prime}_{\nu}(\lambda R)-\frac{(D-2)}{2R}H^{(j)}_{\nu}(\lambda R) \right)-dH^{(j)}_{\nu}(\lambda R)\right]\phi^{\prime}_{\textrm{int}}(\lambda R)\Bigg\}\;.
\end{eqnarray}
It is interesting to point out that by using the properties of the Bessel and Hankel functions with $\rho,x\in\mathbb{R}$ \cite{watson}
\begin{equation}
\overline{J_{\rho}(x)}=J_{\rho}(x)\;,\quad \overline{H^{(1)}_{\rho}(x)}=H^{(2)}_{\rho}(x)\;,
\end{equation}
it is not difficult to prove that 
\begin{equation}
A_{2}(\lambda R)=\overline{A_{1}(\lambda R)}\;,
\end{equation}
and therefore $A_{1}(\lambda R)$ represents the Jost function of the system. 
By imposing the coupled conditions (\ref{bccoupexp}) one does not find a relation that allows for the implicit determination of the 
eigenvalues of the problem. This occurs because the system is defined over an unbounded domain $I_<\cup I_>$ and has, hence, a 
continuous spectrum \cite{bordag1996vacuum}. In order to obtain a discrete spectrum and, consequently, be able to construct the spectral zeta
function associated with the operator $H_{V}$, we enclose the entire system inside an auxiliary large sphere of radius $r=\tilde{R}$ \cite{bordag1996vacuum,fucci2016functional}. Once the boundary conditions at $r=\tilde{R}$ are imposed we obtain an equation that provides implicitly the discrete eigenvalues. The nature of the boundary condition imposed at $r=\tilde{R}$ is not important, since the radius $\tilde{R}$ of
the auxiliary sphere will be sent to infinity once an implicit equation for the eigenvalues is found \cite{bordag1996vacuum}.
By imposing, for simplicity, Dirichlet boundary conditions at $r=\tilde{R}$ on the exterior solution (\ref{extsol}), we obtain the following implicit equation for the eigenvalues
\begin{equation}\label{eqeg}
A_{1}\left(\lambda R\right)H^{(2)}_{\nu}\left(\lambda \tilde{R}\right)+A_{2}\left(\lambda R\right)H^{(1)}_{\nu}\left(\lambda \tilde{R}\right)=0\;,
\end{equation}    
with the coefficients $A_{j}$ given in (\ref{sol}). In the limit $\tilde{R}\to\infty$, the equation (\ref{eqeg}) does not yet provide implicitly the eigenvalues of the system. This is due to the fact that in the limit  $\tilde{R}\to\infty$ one includes a continuous contribution to the spectrum that
originates from the solutions of the radial equation (\ref{app1a}) in absence of the semi-transparent potential at $r=R$ \cite{kirsten2001spectral,bordag1996vacuum}. In order to eliminate the contribution coming from the continuous spectrum,
we impose Dirichlet boundary conditions on the free solution of the radial equation (\ref{app1a}) at $r=\tilde{R}$ \cite{kirsten2001spectral} to
obtain the relation 
\begin{equation}
H^{(2)}_{\nu}\left(\lambda \tilde{R}\right)-H^{(1)}_{\nu}\left(\lambda \tilde{R}\right)=0\;.
\end{equation}
In the limit $\tilde{R}\to\infty$ the last equation generates the continuous spectrum of the system. Therefore, the discrete spectrum associated
with the radial operator (\ref{radoper}) endowed with the boundary conditions (\ref{bccoupexp}) is obtained from the zeros of the following 
characteristic function
\begin{equation}\label{eigeneq}
F_{\nu}(\lambda,R,\tilde{R})=\frac{A_{1}\left(\lambda R\right)H^{(2)}_{\nu}\left(\lambda \tilde{R}\right)+A_{2}\left(\lambda R\right)H^{(1)}_{\nu}\left(\lambda \tilde{R}\right)}{H^{(2)}_{\nu}\left(\lambda \tilde{R}\right)-H^{(1)}_{\nu}\left(\lambda \tilde{R}\right)}\;,
\end{equation} 
in the limit $\tilde{R}\to\infty$.

The characteristic function we have just obtained can be used to construct an integral representation of the spectral zeta function, valid
in the semi-plane $\Re(s)>D/2$, as
\cite{kirsten2001spectral}
\begin{equation}\label{0}
\zeta(s)=\lim_{\tilde{R}\to\infty}\zeta(s,\tilde{R})\;,
\end{equation} 
where 
\begin{eqnarray}\label{1}
\zeta(s,\tilde{R})=\frac{1}{2\pi i}\sum_{l=0}^{\infty}d(l)\int_{\gamma}d\lambda \lambda^{-2s}\frac{\partial}{\partial \lambda}\ln\left[F_{\nu}(\lambda,R,\tilde{R})\right]
\end{eqnarray}
with $\gamma$ denoting a contour that encircles the positive zeros of $F_{\nu}(\lambda,R,\tilde{R})$ in the counterclockwise direction. 
In order to analyze the $\tilde{R}\to\infty$ limit of (\ref{eigeneq}) we first notice that (\cite{olver2010nist}, Section 10.17)
\begin{equation}
H_{\rho}^{(\alpha)}\left(e^{\frac{i\pi\mu}{2}}x\right)\sim x^{-1/2}e^{(-1)^{\alpha}\mu x}\left[1+O\left(\frac{1}{x}\right)\right]\;,
\end{equation} 
for $\alpha=(1,2)$ and $x\in\mathbb{R}$. This relation implies that as $\tilde{R}\to\infty$ we obtain
\begin{eqnarray}\label{2}
F_{\nu}(\lambda,R,\tilde{R})\sim A_{1}\left(\lambda R\right)\;,\quad \Im(\lambda)>0\;,\nonumber\\
F_{\nu}(\lambda,R,\tilde{R})\sim -A_{2}\left(\lambda R\right)\;,\quad \Im(\lambda)<0\;.
\end{eqnarray}
By utilizing the connection formulas (\cite{olver2010nist}, Section 10.27) 
\begin{equation}
H_{\rho}^{(\alpha)}\left(e^{\frac{(-1)^{\alpha-1}i\pi}{2}}\lambda R\right)=(-1)^{\alpha-1}\frac{2}{i\pi}e^{\frac{(-1)^{\alpha}i\pi\rho}{2}}K_{\rho}(\lambda R)\;,
\end{equation}
where $K_{\rho}(z)$ denotes the modified Bessel function of the second kind, and a similar relation for the derivative
\begin{equation}
H_{\rho}^{(\alpha)\prime}\left(e^{\frac{(-1)^{\alpha-1}i\pi}{2}}\lambda R\right)=-\frac{2}{\pi}e^{\frac{(-1)^{\alpha}i\pi\rho}{2}}K'_{\rho}(\lambda R)\;,
\end{equation}
we are able to arrive at an explicit expression for $A_{1}\left(\lambda R\right)$ and $A_{2}\left(\lambda R\right)$ when $\Im(\lambda)>0$ 
and $\Im(\lambda)<0$, respectively, that is 
\begin{eqnarray}\label{coeffA}
A_{j}\left(e^{\frac{(-1)^{j-1}i\pi}{2}}\lambda R\right)=\frac{\lambda R }{2}e^{\frac{(-1)^{j-1}i\pi}{2}}{\cal F}_{\nu}(\lambda R)\;.
\end{eqnarray}
where 
\begin{eqnarray}\label{charact}
{\cal F}_{\nu}(\lambda R)&=&\left[\left(c+\frac{D-2}{2R}a\right)K_{\nu}(\lambda R)-a\lambda K'_{\nu}(\lambda R)\right]I_{\nu}(\lambda R)
\nonumber\\
&+&\left[\left(d+\frac{D-2}{2R}b\right)K_{\nu}(\lambda R)-b\lambda K'_{\nu}(\lambda R)\right]\left(\lambda I'_{\nu}(\lambda R)-\frac{D-2}{2R}I_{\nu}(\lambda R)\right)
\end{eqnarray} 
In order to obtain the last equation we have used the definition of the interior solution given in (\ref{intsol}) and the connection formulas
(\cite{olver2010nist}, Section 10.27)
\begin{equation}
J_{\rho}\left(x e^{\pm\frac{i\pi}{2}}\right)=e^{\pm\frac{i\pi\rho}{2}}I_{\rho}(x)\;,\quad 
J'_{\rho}\left(x e^{\pm\frac{i\pi}{2}}\right)=e^{\pm(\rho-1)\frac{i\pi}{2}}I'_{\rho}(x)\;,
\end{equation}
to find $\phi_{\textrm{int}}(\pm i\lambda r)$ and $\phi'_{\textrm{int}}(\pm i\lambda r)$.
\section{Analytic continuation of the spectral zeta function}

As we have previously mentioned, 
the integral representation of the spectral zeta function in (\ref{1}) is valid in the region $\Re(s)>D/2$. The first step 
of the analytic continuation consists in deforming the integration contour $\gamma$ in (\ref{1}) along the imaginary axis.   
After performing the contour deformation and, in the process, utilizing the relation
\begin{equation}\label{equal}
\frac{\partial}{\partial \lambda}\ln A_{1}\left(e^{\frac{i\pi}{2}}\lambda R\right)=\frac{\partial}{\partial \lambda}\ln A_{2}\left(e^{-\frac{i\pi}{2}}\lambda R\right)\;,
\end{equation}    
that can be directly proved from (\ref{coeffA}), we obtain an integral representation valid in the strip $1/2<\Re(s)<1$ \cite{kirsten2001spectral}
\begin{equation}\label{zetafin}
\zeta(s)=\sum_{l=0}^{\infty}d(l)\zeta_{\nu}(s)\;,
\end{equation} 
with
\begin{equation}\label{zetafinl}
\zeta_{\nu}(s)=\frac{\sin(\pi s)}{\pi}\int_{0}^{\infty}d\lambda \lambda^{-2s}\frac{\partial}{\partial \lambda}\ln{\cal F}_{\nu}(\lambda R)\;.
\end{equation}
It is important to point out that the integral representation (\ref{zetafin}) has been obtained under the assumption that $\lambda=0$ is not an eigenvalue of the system. By using the 
small argument expansion of the modified Bessel functions and their first derivative, it is not very difficult to show that 
as $\lambda\to 0$
\begin{equation}\label{as}
{\cal F}_{\nu}(\lambda R)=\frac{1}{2\nu}\left[\left(c+\frac{l}{R}d\right)+\frac{D-2+l}{R}\left(a+\frac{l}{R}b\right)\right]+O(\lambda)\;.
\end{equation}
This implies that $\lambda=0$ is not an eigenvalue as long as the $O(1)$ term in (\ref{as}) does not vanish. We would like add
that even if $\lambda=0$ were an eigenvalue of the system, one would still obtain an integral representation of the form (\ref{zetafin}) for the 
spectral zeta function, albeit with the characteristic function $\lambda^{-\alpha}{\cal F}_{\nu}(\lambda R)$ where $\alpha$ coincides with 
the multiplicity of the eigenvalue $\lambda=0$ \cite{kirsten2001spectral}.  

The second step in the process of analytic continuation of $\zeta(s)$ consists in the subtraction, and subsequent addition, of a suitable
number of terms of the asymptotic expansion of the derivative in the integrand of (\ref{zetafinl}). 
In order to make sure that the convergence of the summation over the angular momenta
improves, we need an asymptotic expansion of the logarithm of the characteristic function for large $\nu$ and uniform in the variable $z=\lambda R/\nu$. 
After performing the change of variables $\lambda R=\nu z$ in (\ref{zetafinl}) we obtain
\begin{equation}\label{zetal1}
\zeta_{\nu}(s)=\frac{\sin(\pi s)}{\pi}\left(\frac{R}{\nu}\right)^{2s}\int_{0}^{\infty}dz z^{-2s}\frac{\partial}{\partial z}\ln{\cal F}_{\nu}(\nu z)\;,
\end{equation}  
where, by rearranging the expression (\ref{charact}) in a form more suitable for an asymptotic expansion, we have 
\begin{equation}\label{charact1}
{\cal F}_{\nu}(\nu z)=\delta+\alpha K_{\nu}(\nu z)I_{\nu}(\nu z)+\gamma \frac{\nu z}{R}\left[K_{\nu}(\nu z)I_{\nu}(\nu z)\right]'-b\frac{\nu^{2}z^{2}}{R^{2}}K'_{\nu}(\nu z)I'_{\nu}(\nu z)\;,
\end{equation}
with the coefficients
\begin{equation}\label{coeffi}
\delta=\frac{a+d}{2R}\;,\quad \alpha=\frac{2R[(D-2)(a-d)+2Rc]-(D-2)^{2}b}{4R^{2}}\;,\quad \gamma=\frac{(D-2)b-(a-d)R}{2R}\;. 
\end{equation}

By subtracting, and then adding, in the integrand of (\ref{zetal1}), the first $N$ terms of the uniform asymptotic expansion 
of $\ln{\cal F}_{\nu}(\nu z)$, derived in Appendix \ref{app1}, one obtains an expression for the spectral zeta function in (\ref{zetafin}) which 
is valid in the semi-plane $\Re(s)>(D-N-2)/2$. Since the form of the uniform asymptotic expansion 
of $\ln{\cal F}_{\nu}(\nu z)$ depends on whether $b\neq 0$ or $b=0$ in (\ref{charact1}) it is convenient to treat the two cases separately. 
\paragraph{Case $b\neq 0$.}
In this case, by using the uniform asymptotic expansion (\ref{asymcharb}) one obtains the expression 
\begin{equation}\label{zetab}
\zeta_{b,D}(s)=Z(s)+\sum_{i=0}^{N}A_{i}(s)\;,
\end{equation} 
where 
\begin{equation}
Z(s)=\frac{\sin(\pi s)}{\pi}R^{2s}\sum_{l=0}^{\infty}d(l)\nu^{-2s}\int_{0}^{\infty}dz z^{-2s}\frac{\partial}{\partial z}\left\{\ln{\cal F}_{\nu}(\nu z)+\ln t-\sum_{n=1}^{N}\frac{\Omega_{n}(t)}{\nu^{n}}\right\}\;,
\end{equation}
and the functions $A_{i}(s)$ are computed by explicitly integrating the asymptotic terms and by then summing over the angular momenta, i.e. 
\begin{equation}
A_{0}(s)=\frac{R^{2s}}{2}S_{D}(s)\;,
\end{equation} 
and, for $i\geq 1$,
\begin{equation}\label{Ai}
A_{i}(s)=-\frac{R^{2s}}{\Gamma(s)}S_{D}\left(s+\frac{i}{2}\right)\sum_{l=0}^{2i}\omega_{i,l+i}\frac{\Gamma\left(s+\frac{l+i}{2}\right)}{\Gamma\left(\frac{l+i}{2}\right)}\;,
\end{equation} 
where $S_{D}(s)$ represents the zeta function on the surface of the sphere,
\begin{equation}
S_{D}(s)=\sum_{l=0}^{\infty}d(l)\nu^{-2s}=\sum_{l=0}^{\infty}d(l)\left(l+\frac{D-2}{2}\right)^{-2s}\;.
\end{equation}
We would like to point out that by using the expression for the degeneracy $d(l)$ provided in (\ref{degeneracy}), the zeta function 
$S_{D}(s)$ can be written as a combination of two Barnes zeta functions $\zeta_{B}(s,a,b)$,
\begin{equation}
S_{D}(s)=\zeta_{B}\left(2s,\frac{D}{2},D-1\right)+\zeta_{B}\left(2s,\frac{D}{2}-1,D-1\right)\;,
\end{equation}
where $\zeta_{B}(s,a,b)$ can, in turn, be written, when $b\in\mathbb{N}^{+}$, in terms of a sum of the, perhaps more familiar, Hurwitz zeta function as follows
\begin{equation}
\zeta_{B}(s,a,b)=\sum_{k=1}^{b}\frac{(-1)^{b-k}B_{b-k}^{(b)}(a)}{(k-1)!(b-k)!}\zeta_{H}(s-k+1,a)\;,
\end{equation}
with $B_{m}^{(n)}(x)$ denoting the generalized Bernoulli polynomials defined in \cite{olver2010nist}, Section 24.16. In more detail, one obtains
\begin{eqnarray}\label{essd}
S_{D}(s)&=&-\left(\frac{D}{2}-1\right)^{-2s-1}\sum_{k=1}^{D-1}\frac{(-1)^{D-1-k}}{(k-1)!(D-1-k)!}B_{D-1-k}^{(D-1)}\left(\frac{D}{2}\right)\left(\frac{D}{2}-1\right)^{k}\nonumber\\
&+&\sum_{k=1}^{D-1}\frac{(-1)^{D-1-k}[(-1)^{D-1-k}+1]}{(k-1)!(D-1-k)!}B_{D-1-k}^{(D-1)}\left(\frac{D}{2}-1\right)
\zeta_{H}\left(2s-k+1;\frac{D}{2}-1\right)\;,
\end{eqnarray} 
where the last expression has been obtained by utilizing the property \cite{srivastava88} $B_{n}^{(\alpha)}(x)=(-1)^{n}B_{n}^{(\alpha)}(\alpha-x)$  
valid for arbitrary real or complex parameters $\alpha$.

\paragraph{Case $b=0$.}
By exploiting the uniform asymptotic expansion (\ref{asymchar0}) we obtain, in this case, the following expression for the 
spectra zeta function
\begin{equation}\label{zetab0}
\zeta_{0,D}(s)=\tilde{Z}(s)+\sum_{i=1}^{N}\tilde{A}_{i}(s)\;,
\end{equation} 
where 
\begin{equation}
\tilde{Z}(s)=\frac{\sin(\pi s)}{\pi}R^{2s}\sum_{l=0}^{\infty}d(l)\nu^{-2s}\int_{0}^{\infty}dz z^{-2s}\frac{\partial}{\partial z}\left\{\ln{\cal F}_{\nu}(\nu z)-\sum_{n=1}^{N}\frac{\tilde{\Omega}_{n}(t)}{\nu^{n}}\right\}\;,
\end{equation}
and the functions $A_{i}(s)$, calculated in the same way as in the case $b\neq 0$, have the form, with $i\geq 1$,
\begin{equation}\label{Atildei}
\tilde{A}_{i}(s)=-\frac{R^{2s}}{\Gamma(s)}S_{D}\left(s+\frac{i}{2}\right)\sum_{l=0}^{2i}\tilde{\omega}_{i,l+i}\frac{\Gamma\left(s+\frac{l+i}{2}\right)}{\Gamma\left(\frac{l+i}{2}\right)}\;.
\end{equation}  

With the explicit analytically continued expressions for the spectral zeta function in (\ref{zetab}) and (\ref{zetab0}) we can compute the Casimir 
energy of the scalar field propagating under the influence of the shell potential.

\section{Casimir energy}
\label{casEner}

Within the framework of the spectral zeta function regularization, the Casimir energy of the scalar field is defined as follows \cite{bordag2009advances,kirsten2001spectral}: Introduce the function 
\begin{equation}\label{eq:Zeta1}
	{\cal E}(\varepsilon)=\frac{{\mu}^{2\varepsilon}}{2} \zeta\left(\varepsilon-\frac{1}{2}\right)\;,
\end{equation}
where $\mu$ is a parameter with dimensions of mass introduced to maintain the correct dimensions, and $\hbar=c=1$.
The Casimir energy of the system is defined as the limit, {\it if} it exists, 
\begin{equation}\label{casenergy}
E_{\textrm{Cas}}=\lim_{\varepsilon\to 0}{\cal E}(\varepsilon)\;.
\end{equation}
We notice, from the analytically continued expressions obtained in the previous section, that the spectral zeta function generally develops a simple pole at the point $s=-1/2$, and hence, the limit (\ref{casenergy}) does not always exist. In fact, in a neighborhood of $\varepsilon=0$
the function (\ref{eq:Zeta1}) has the expansion
\begin{equation}\label{cale}
{\cal E}(\varepsilon)=\frac{1}{2\varepsilon}\textrm{Res}\,\zeta\left(-\frac{1}{2}\right)
+\frac{1}{2}\left[\textrm{FP}\,\zeta\left(-\frac{1}{2}\right)+\textrm{Res}\,\zeta\left(-\frac{1}{2}\right)\ln\mu^2\right]+O(\varepsilon)\;.
\end{equation} 
It is clear, from this expression, that the limit in (\ref{casenergy}), and hence the Casimir energy of the system, is well-defined if
the residue of the spectral zeta function at the point $s=-1/2$ vanishes identically. 
We would like to mention that according to the general theory of the spectral zeta function, the residue at $s=-1/2$ in $D$ dimensions is proportional to the heat kernel coefficient $a_{(D+1)/2}$ \cite{kirsten2001spectral}. This implies that a configuration for which the 
heat kernel coefficient $a_{(D+1)/2}=0$ is guaranteed to have a well-defined Casimir energy (which does not need further renormalization) \cite{bordag2009advances}.  

We, therefore, focus on those shell-potentials for which the residue of the zeta function at $s=-1/2$ vanishes. In order to compute the zeta function of the system at $s=-1/2$ we only need to consider, in (\ref{zetab}) and in (\ref{zetab0}), $N=D$ terms of the uniform asymptotic expansion of the relevant characteristic function. Hence, we assume, from now on, that $N=D$.
The residue of the spectral zeta function at $s=-1/2$ depends explicitly on the meromorphic structure of the function $S_{D}(s)$ and the 
Gamma functions that define the terms $A_{i}(s)$ and $\tilde{A}_{i}(s)$. Obviously, no contribution to the residue can come from $Z(s)$ 
and $\tilde{Z}(s)$ since they are analytic for $\Re(s)>-1$ by construction.  
To analyze the pole structure of $S_{D}(s)$, we focus on its expression in terms of the Hurwitz zeta function in (\ref{essd}). 
It is not difficult to realize that $S_{D}(s)$ has no poles for $\Re(s)<0$. The only possible poles are simple and occur at $s=m/2$ with $m=1,\ldots, D-1$. In particular, one can show that 
\begin{equation}\label{residuemover2}
\textrm{Res}\,S_{D}\left(\frac{m}{2}\right)=\frac{(-1)^{D-m-1}[(-1)^{D-m-1}+1]}{2(m-1)!(D-m-1)!}B_{D-m-1}^{(D-1)}\left(\frac{D}{2}-1\right)\;.
\end{equation}  
Now, for $m=1$ the above residue vanishes. In fact, when $D$ is odd, this statement is obviously true, while when $D$ is even we can arrive at the same conclusion by using the fact, proved in Appendix \ref{app2}, that $B_{n-1}^{(n)}(j)=0$ for $j=1,\ldots,n-1$. We can, hence, say that $\textrm{Res}\,S_{D}\left(1/2\right)=0$.  

These last remarks imply that the terms $A_{i}$ and $\tilde{A}_{i}$, in (\ref{Ai}) and (\ref{Atildei}), respectively, with $3\leq i\leq D$ contribute to the residue of the spectral zeta function at $s=-1/2$. In addition, the function $\Gamma[(-1+l+i)/2]$ in (\ref{Ai}) and (\ref{Atildei}) develops a simple pole for $l=0$ when $i=1$. With all the terms that contribute to the residue at $s=-1/2$ accounted for, we can finally write, from (\ref{Ai}), that, for $b\neq 0$,
\begin{eqnarray}\label{resbno0}
\textrm{Res}\,\zeta_{b,D}\left(-\frac{1}{2}\right)=\frac{1}{2\sqrt{\pi}R}\Bigg\{\sum_{i=0}^{D-3}\textrm{Res}\,S_{D}\left(\frac{i+2}{2}\right)\sum_{l=0}^{2i+6}\omega_{i+3,l+i+3}\frac{\Gamma\left(\frac{i+l+2}{2}\right)}{\Gamma\left(\frac{i+l+3}{2}\right)}+
S_{D}(0)\frac{2\delta R^{2}}{\sqrt{\pi}b}\Bigg\}\;.
\end{eqnarray}   
Similarly, for $b=0$, by using (\ref{Atildei}), one obtains
\begin{eqnarray}\label{resb0}
\textrm{Res}\,\zeta_{0,D}\left(-\frac{1}{2}\right)=\frac{1}{2\sqrt{\pi}R}\Bigg\{\sum_{i=0}^{D-3}\textrm{Res}\,S_{D}\left(\frac{i+2}{2}\right)\sum_{l=0}^{2i+6}\tilde{\omega}_{i+3,l+i+3}\frac{\Gamma\left(\frac{i+l+2}{2}\right)}{\Gamma\left(\frac{i+l+3}{2}\right)}+
S_{D}(0)\frac{R\tilde{\alpha}-\tilde{\gamma}}{2\sqrt{\pi}R}\Bigg\}\;.
\end{eqnarray} 

To complete the computation of the Casimir energy, we need to evaluate the finite part of the spectral zeta function at the point $s=-1/2$. 
From (\ref{zetab}) it is not difficult to obtain 
\begin{equation}
\textrm{FP}\,\zeta_{b,D}\left(-\frac{1}{2}\right)=Z\left(-\frac{1}{2}\right)+\sum_{i=1}^{D}\textrm{FP}A_{i}\left(-\frac{1}{2}\right)\;.
\end{equation}
The explicit expression of the finite part of the terms $A_{i}(s)$ are found to be
\begin{equation}
\textrm{FP}A_{1}\left(-\frac{1}{2}\right)=-\frac{2R\delta}{\pi b}\left[\left(1-\ln(2R)\right)S_{D}(0)-\frac{S'_{D}(0)}{2}\right]\;,
\end{equation}
\begin{equation}
\textrm{FP}A_{2}\left(-\frac{1}{2}\right)=-\frac{\left(5 b^2-64 \alpha  b R^2+32 b \gamma  R+128 \delta ^2 R^4\right)}{128 b^2 R}S_{D}\left(\frac{1}{2}\right)\;,
\end{equation}
and, for $3\leq i\leq D$, we have 
\begin{eqnarray}
\textrm{FP}A_{i}\left(-\frac{1}{2}\right)&=&\frac{1}{2\sqrt{\pi} R}\sum_{l=0}^{2i}\omega_{i,l+i}\frac{\Gamma\left(\frac{l+i-1}{2}\right)}{\Gamma\left(\frac{l+i}{2}\right)}\Bigg[\textrm{FP}S_{D}\left(\frac{i-1}{2}\right)\nonumber\\
&+&\textrm{Res}S_{D}\left(\frac{i-1}{2}\right)\left(\gamma_{E}-2+2\ln(2R)+\Psi\left(\frac{l+i-1}{2}\right)\right)\Bigg]\;,
\end{eqnarray}
with $\gamma_{E}$ denoting the Euler's constant.
The explicit expression for $\textrm{Res}S_{D}(m/2)$ with $m=\{1,\ldots, D-1\}$ is displayed in (\ref{residuemover2}), while the finite part is
\begin{eqnarray}
\textrm{FP}S_{D}\left(\frac{m}{2}\right)&=&-\sum_{k=1}^{D-1}\frac{(-1)^{D-1-k}}{(k-1)!(D-1-k)!}B_{D-1-k}^{(D-1)}\left(\frac{D}{2}\right)\left(\frac{D}{2}-1\right)^{k-m-1}\nonumber\\
&+&2\sum_{{k=1}\atop {k\neq m}}^{D-1}\textrm{Res}S_{D}\left(\frac{k}{2}\right)\zeta_{H}\left(m-k+1;\frac{D}{2}-1\right)-2\textrm{Res}S_{D}\left(\frac{m}{2}\right)\Psi\left(\frac{D}{2}-1\right)\;.
\end{eqnarray} 

For the case $b=0$ we similarly have
\begin{equation}
\textrm{FP}\,\zeta_{0,D}\left(-\frac{1}{2}\right)=\tilde{Z}\left(-\frac{1}{2}\right)+\sum_{i=1}^{D}\textrm{FP}\tilde{A}_{i}\left(-\frac{1}{2}\right)\;.
\end{equation}
The finite part of the functions $\tilde{A}_{i}(s)$ at $s=-1/2$ is 
\begin{equation}
\textrm{FP}\tilde{A}_{1}\left(-\frac{1}{2}\right)=\frac{\tilde{\gamma}S_{D}(0)}{4\pi R^{2}}+\frac{R\tilde{\alpha}
-\tilde{\gamma}}{4\pi R^{2}}\left[(-1+\ln(2R))S_{D}(0)+S'_{D}(0)\right]\;,
\end{equation}
\begin{equation}
\textrm{FP}\tilde{A}_{2}\left(-\frac{1}{2}\right)=-\frac{\left(3 \tilde{\gamma }^2+8 R^2 \tilde{\alpha }^2-8 R \tilde{\alpha } \tilde{\gamma }\right)}{128 R^3}S_{D}\left(\frac{1}{2}\right)\;, 
\end{equation}
and, for $3\leq i\leq D$, we have 
\begin{eqnarray}
\textrm{FP}\tilde{A}_{i}\left(-\frac{1}{2}\right)&=&\frac{1}{2\sqrt{\pi} R}\sum_{l=0}^{2i}\tilde{\omega}_{i,l+i}\frac{\Gamma\left(\frac{l+i-1}{2}\right)}{\Gamma\left(\frac{l+i}{2}\right)}\Bigg[\textrm{FP}S_{D}\left(\frac{i-1}{2}\right)\nonumber\\
&+&\textrm{Res}S_{D}\left(\frac{i-1}{2}\right)\left(\gamma_{E}-2+2\ln(2R)+\Psi\left(\frac{l+i-1}{2}\right)\right)\Bigg]\;.
\end{eqnarray}

The next step consists in determining which shell potentials lead to a vanishing residue, and hence a well-defined Casimir energy. Unfortunately, due to the complicated form of the expressions (\ref{resbno0}) and (\ref{resb0}), we are not able to identify in full generality the shell potentials characterized by a zero residue. However, we can analyze particular cases in which one obtains a well-defined Casimir energy. 
We will focus on the case $D=3$ since it is the one that is most important in applications. In this particular case the residues in (\ref{resbno0}) and (\ref{resb0}) simplify to, respectively,
\begin{equation}\label{res1bno0}
\textrm{Res}\,\zeta_{b,3}\left(-\frac{1}{2}\right)=\frac{(a+d) \left[5\left((a-d)^2+ad+3\right) R^2 -10 b(a-d)R+7 b^2\right]}{15 \pi  b^3}\;,
\end{equation}
and
\begin{equation}\label{res1b0}
\textrm{Res}\,\zeta_{0,3}\left(-\frac{1}{2}\right)=\frac{-70 c^3 d^3 R^3+140 c^2 d^2 \left(d^2-1\right) R^2-98 c d \left(d^2-1\right)^2 R+24 \left(d^2-1\right)^3}{105 \sqrt{\pi } \left(d^2+1\right)^3 R}\;,
\end{equation} 
where we have used (\ref{coeffi}) and the fact that $ad-bc=1$ to express them in terms of the parameters $(a,b,c,d)$ . 

We are now in a position to discuss the circumstances in which these residues vanish. Before proceeding, we would like to point out that
for the $\delta$-shell potential (namely $a=d=1$ and $b=0$) in $D=3$ the result (\ref{res1b0}) clearly shows that 
the residue of the spectral zeta function does not vanish. This implies that for the $\delta$-shell potential further renormalization of the Casimir energy is necessary (see e.g. \cite{bordag1999ground} and references therein).      

\paragraph{Case $b\neq 0$.}
In this case the residue in (\ref{res1bno0}) vanishes identically for the shell potentials characterized by $a+d=0$. When $a+d\neq 0$, instead,
the residue does not vanish identically but we can analyze the numerator of (\ref{res1bno0}) to determine whether values of the parameters  $a,b,d$, and $R$ exist for which the residue is zero. For convenience, let us define
\begin{equation}
f(a,b,d,R)=5\left((a-d)^2+ad+3\right) R^2 -10 b(a-d)R+7 b^2\;.
\end{equation}
We can demonstrate that $f(a,b,d,R)>0$ in the region $D=\{(a,b,d,R)\in\mathbb{R}^{4} | (a,d)\in\mathbb{R}^{2}, b\in\mathbb{R}-\{0\}, R>0\}$.
We notice that $f(a,b,d,R)$ can be written as a polynomial in $R$ of degree two. For the coefficient of $R^{2}$ one can prove that 
$5[(a-d)^2+ad+3]>0$ for all $(a,b,d)\in\mathbb{R}^{3}$. Moreover, the discriminant of the polynomial satisfies $\Delta=-20 b^2 \left(2 a^2+3 a d+2 d^2+21\right)<0$ for all $(a,d)\in\mathbb{R}^{2}$. These observations imply that $f(a,b,d,R)>0$ for all $(a,b,d,R)\in\mathbb{R}^{4}$. In particular, we can conclude that $f(a,b,d,R)>0$ in the region $D$.
The last argument shows that the residue of the spectral zeta function in the case $b\neq 0$ is vanishing only for shell potentials described by the condition $a+d=0$.

We can, therefore, compute the Casimir energy under this condition. For $b\neq 0$ and for $D=3$ the relevant finite parts read
\begin{equation}\label{finitepart1}
\textrm{FP}A_{1}\left(-\frac{1}{2}\right)=\frac{(a+d) [12 \log (A)+\log (R)-2]}{12 \pi  b}\;,
\end{equation} 
where $A$ is the Glaisher–Kinkelin constant,
\begin{equation}
\textrm{FP}A_{2}\left(-\frac{1}{2}\right)=0\;,
\end{equation}
and
\begin{eqnarray}\label{finitepart3}
\textrm{FP}A_{3}\left(-\frac{1}{2}\right)&=&\frac{(a+d)}{60 \pi  b^3}\Big\{\log (2 R) \left[40 R^2 \left((a+d)^2-3 b c\right)+80 b R (d-a)+51 b^2\right]\nonumber\\
&-&20 b R \left[a (4 \gamma -5+8\log (2))+6 c R (\gamma -1+2\log (2))+d (-4 \gamma +5-8 \log (2))\right])\nonumber\\
&+&40 R^2 \left[\gamma -1+2\log (2)) (a+d)^2+b^2 (51 \gamma -79+102 \log (2)\right]\Big\}
\end{eqnarray}
In the case of vanishing residue, $d=-a$ and $c=-(1+a^2)/b$ (which can be easily obtained by noticing that $ad-bc=1$ and $d=-a$) which immediately  shows that also the finite part of the spectral zeta function at $s=-1/2$, given in (\ref{finitepart1})-(\ref{finitepart3}), vanishes identically.
This implies that only $Z(-1/2)$ contributes to the Casimir energy, that is
\begin{equation}\label{Energy}
E_{\textrm{Cas}}=\frac{1}{2}Z\left(-\frac{1}{2}\right)\;,\quad\textrm{with}\quad d=-a,\; c=-\frac{1+a^2}{b}\;.
\end{equation}
The Casimir energy (\ref{Energy}) depends on the parameters $a$, $b$ and the radius $R$ and can only be computed numerically due to the complexity of the integral that defines the function $Z(-1/2)$. 

We would like to make an important remark at this point. In order to have a well-defined Casimir energy we need to consider shell potentials 
associated with self-adjoint extensions that generate only positive eigenvalues. The presence of negative eigenvalues signifies that the quantum theory under consideration is non-unitary which would lead to phenomena such as vacuum decay (see e.g. \cite{coleman}). If a two-interval self-adjoint extension of (\ref{radoper}) develops negative eigenvalues, they will appear as positive zeroes of the corresponding characteristic function ${\cal F}_{\nu}(\nu z)$ in (\ref{charact1}). An attempt at determining whether positive zeroes exist for ${\cal F}_{\nu}(\nu z)$ in (\ref{charact1}) for all allowed values of the parameters $\alpha$, $\gamma$ and $b$ 
has proved to be surprisingly difficult. While we know that ${\cal F}_{\nu}(\nu z)$ tends to a constant value as $z\to 0$, as shown in (\ref{as}), and that 
\begin{equation}
{\cal F}_{\nu}(\nu z)\sim \frac{b\nu z}{2R^2}\;,\;\textrm{as}\; z\to+\infty\;,
\end{equation}         
we cannot provide a proof that demonstrates whether or not ${\cal F}_{\nu}(\nu z)$ is monotonic in the interval $(0,\infty)$. In fact, the analysis of the monotonicity 
properties of products and ratios of modified Bessel functions and their derivatives is currently quite an active field of research (see e.g. \cite{baricz11,segura21}). Nevertheless, numerical experiments show that ${\cal F}_{\nu}(\nu z)$ is indeed a monotonic function that is increasing when $b>0$ and decreasing when $b<0$. This observation, together with the $z\to 0$ behavior displayed in (\ref{as}), shows that a single positive zero of ${\cal F}_{\nu}(\nu z)$ exists when the parameter $b$ and the constant given in (\ref{as}) have the {\it opposite} sign. Since we want to avoid self-adjoint extensions with negative eigenvalues, we restrict the analysis of the Casimir energy to those potential configurations for which $b$ and the constant in (\ref{as}) have the same sign. To determine the range of values of $a$, $b$ for which this occurs,
we rewrite the leading small-$z$ term shown in (\ref{as}), specializing it to the case under consideration, as
\begin{equation}\label{constant}
{\cal G}_{l}(a,b,R)=\frac{1}{(2l+1)b R^2}\left[l(l+1)b^2+abR-(a^{2}+1)R^2\right]\;.
\end{equation} 

Let us consider the case $b>0$. Denote the roots of the numerator of (\ref{constant}) as a function of $b$ as
\begin{eqnarray}
b_{0}&=&\frac{(a^2+1)R}{a}\;,\;\textrm{when} \;l=0\;,\\
b^{\pm}_{l}&=&\frac{R}{2l(l+1)}\left[\pm\sqrt{a^{2}(2l+1)^{2}+4l(l+1)}-a\right]\;,\;\textrm{when} \;l\geq 1\;.
\end{eqnarray} 

The leading term of ${\cal F}_{\nu(l)}(\nu(l) z)$ shown in (\ref{constant}) is positive if $b>b_{0}$, when $l=0$, and $b>b^{+}_{l}$ for $l\geq 1$.
In order for both $b$ and ${\cal G}_{l}(a,b,R)$ to be positive we must have $b_{0}>0$ and $b^{+}_{l}>0$, which occurs when $a>0$.   
Since the magnitude of $b^{+}_{l}$ decreases as the value of $l$ increases and since $b_{0}>b^{+}_{l}$, for $l\geq 1$, (which can be shown by noticing that ${\cal G}_{l}(a,b_{0},R)>0$ for $l\geq 1$, $R>0$, and $a>0$), we can, hence, conclude that both $b$ and ${\cal G}_{l}(a,b,R)$ are positive if $b>(a^2+1)R/a$ and $a>0$.  

Let us turn our attention, now, to the case $b<0$. The function (\ref{constant}) is also negative if $a<0$ and $b<b_{0}$, when $l=0$, 
and $b<b^{-}_{l}$ when $l\geq 1$. Similarly to the case above, $b^{-}_{l}$ decreases in magnitude as $l$ increases which implies that 
$b_{0}<b^{-}_{l}$ for $l\geq 1$. These observations allow us to conclude that both $b$ and ${\cal G}_{l}(a,b,R)$ are negative if $b<(a^2+1)R/a$ and $a<0$.

\begin{figure}[h!]
	\centering
	\includegraphics[width=.65\textwidth]{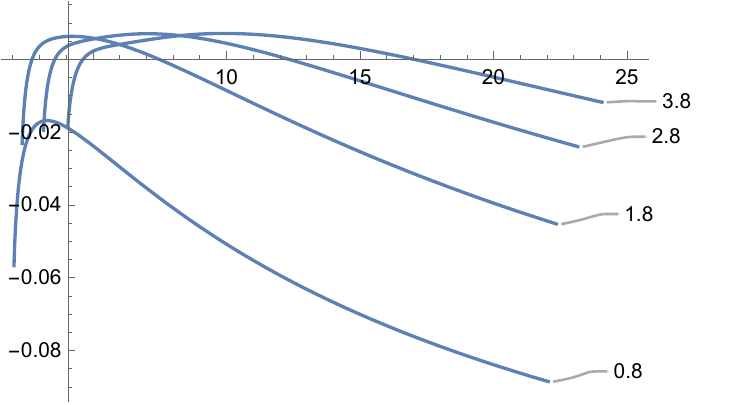}
	\caption{Plot of the Casimir energy $E_{\textrm{Cas}}$ for $R=1$ as a function of the parameter $a$ when $b>(a^2+1)R/a$. The graphs are labeled according to the value of the parameter $b$ used for evaluation of the energy.}   
	\label{fig:1}
\end{figure}

In Figure \ref{fig:1} we show the Casimir energy for $R=1$ as a function of $a>0$ when $b$ varies in the region $b>(a^2+1)R/a$. The graphs displayed there are obtained for different values of $b$ and are labeled accordingly. Based on the graphs in Figure \ref{fig:1} 
we can offer a few comments. We notice that when $a>0$ there are values of $b$ for 
which the energy is always negative. However, as $b$ increases, the energy increases as well. When $b$ becomes large enough, the Casimir energy becomes positive for a finite range of values of $a$. In this situation, particular values of $a$ and $b$ exist for which 
the Casimir energy vanishes identically. Furthermore, we would like to point out that, for a fixed $b$, there exists a value of $a$ for which the energy reaches a maximum. This value of $a$ appears to be an increasing function of $b$.    

The existence of shell-potentials generating a vanishing Casimir energy is an interesting feature of this model since they 
mimic a configuration in which no potential is present. In fact, the self-adjoint extension characterized by $a=d=1$ and $b=c=0$ in (\ref{bccoupexp}) is associated with a system with no potential and has, consequently, an identically vanishing Casimir energy. 
This implies that, for the purpose of evaluating the Casimir energy, configurations with these shell potentials are equivalent to those without 
any potential.   

\paragraph{Case $b=0$.}
This case coincides with the $\delta$-$\delta'$ potential which has been considered in \cite{romaniega2022casimir}. 
Here, we have to analyze the numerator in (\ref{res1b0}) in order to discern whether or not a range for the parameters 
$(c,d,R)$ exists such that the residue vanishes. To this end, we notice that the numerator of (\ref{res1b0}) is a polynomial of degree three in the variable $cR$. It is not very difficult to show that this polynomial is monotone since its $cR$-derivative never vanishes. This implies, then, that there exists exactly one value of $cR$ for which the residue vanishes. This value can be shown to be
\begin{equation}\label{equationzero}
cR=\frac{\left(d^2-1\right)}{105 d}\left[70-\sqrt[3]{245 \left(3 \sqrt{30}-5\right)}+ \sqrt[3]{245 \left(3 \sqrt{30}+5\right)}\right] \;.
\end{equation}
We can, therefore, conclude that in the case $b=0$, for any given positive radius $R$ we can find a shell potential characterized by a pair 
of values $(c,d)$ that satisfies (\ref{equationzero}) and, hence, leads to a vanishing residue of the spectral zeta function.

\begin{figure}[h!]
	\centering
	\includegraphics[width=.65\textwidth]{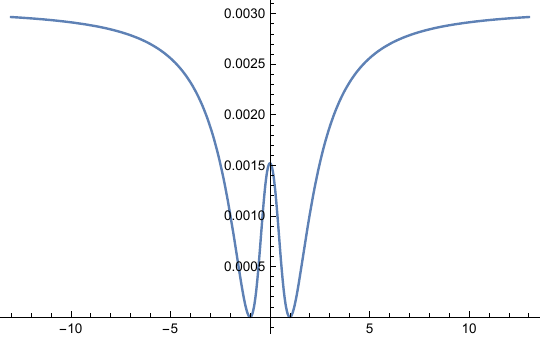}
	\caption{Plot of the Casimir energy $E_{\textrm{Cas}}$ for $R=1$ as a function of the parameter $d$ when $c$ satisfies (\ref{equationzero}).}   
	\label{fig:2}
\end{figure}

In Figure \ref{fig:2} we show a plot of the Casimir energy for $R=1$ as a function of the parameter $d$ when the parameter $c$ satisfies the constraint found in (\ref{equationzero}). As it is clear from the graph, the energy is always positive as it has already been observed in \cite{romaniega2022casimir}.  

The plot presented in \cite{romaniega2022casimir} is different then the one obtained in Figure \ref{fig:2} simply because they chose to evaluate the Casimir energy with respect to a parameter that is different from ours. Nevertheless, some important limiting cases can be directly compared.
The limit $d\to 0$ corresponds to a shell potentials that imposes Dirichlet boundary conditions on $R_{-}$ and Robin boundary conditions on 
$R_{+}$ and the limit $d\to\pm \infty$, corresponds, instead, to a shell potential characterized by Robin boundary conditions on 
$R_{-}$ and Dirichlet boundary conditions on $R_{+}$ \cite{romaniega2022casimir,albeverio2000singular}. Figure \ref{fig:2} shows that in these limits the value of the Casimir energy matches the ones found in \cite{romaniega2022casimir}.  
We also point out that for $d=\pm 1$ the Casimir energy vanishes since, in this case, there is no potential (the boundary conditions (\ref{bccoupexp}) reduce to the identity). 

We would like to take the opportunity, at this point, to make a final comment for the case $b=0$. As we have mentioned in this section, the residue of the spectral zeta function at $s=-1/2$ is proportional to the heat kernel coefficient $a_{(D+1)/2}$. This was proved, for instance, in the case of point potentials corresponding to the case $b=0$ in \cite{1999heat}. The coefficients of the asymptotic expansion of the heat kernel for point potentials were computed, by utilizing the multiple reflection expansion formalism, in \cite{bordag2001multiple}. 
By using the expression (\ref{resb0}) and the relation \cite{kirsten2001spectral}
\begin{equation}\label{HKC1}
a_{\frac{D}{2}-s}=(4\pi)^{D/2}\Gamma(s)\textrm{Res}\,\zeta(s)\;,
\end{equation}
which provides the coefficients up to $a_{(D-1)/2}$, we can explicitly evaluate the first few coefficients for point potentials when $b=0$. 
In particular, for $D=3$, the expression (\ref{res1b0}), rewritten in terms of the original parameters $\tilde{\alpha}$ and $\tilde{\gamma}$, allows us to obtain, using (\ref{HKC1}), 
\begin{equation}\label{a2}
a_{2}=\frac{2 \pi }{105 R^4} \left(5 \tilde{\gamma }^3-35 R^3 \tilde{\alpha }^3+35 R^2 \tilde{\alpha }^2 \tilde{\gamma }-21 R \tilde{\alpha } \tilde{\gamma }^2\right)\;.
\end{equation}  
A quick comparison of this coefficient with the one provided in Eq. (49) of \cite{bordag2001multiple} shows a discrepancy between our coefficient  
and theirs. A careful dimensional analysis of the parameters $(a,b,c,d)$ describing the matching conditions shows that the terms 
$\alpha$ and $\gamma$ used in \cite{bordag2001multiple} have the dimensions of a pure number and of a length, respectively (when $\hbar=c=1$).
The particular combination of $\alpha$ and $\gamma$ which used to construct the expression for $a_2$ in Eq. (49) of \cite{bordag2001multiple} does not have the right dimensions given that the product $R a_2$ must be dimensionless \cite{bordag2009advances,kirsten2001spectral}. This problem, unfortunately, affects also the other heat kernel coefficients presented there.
The discrepancy can be traced back to the presence of incorrect coefficients in the characteristic equation used in Eq. (47) of \cite{bordag2001multiple} to express the spectral zeta function. The coefficients to use for the characteristic equation are, instead, the ones presented in (\ref{coeffi}). When the correct characteristic equation is used in \cite{bordag2001multiple} the discrepancy should obviously disappear.     

\section{Conclusions}

In this work we have focused on the analysis of the Casimir energy of a scalar field under the influence of a general spherically symmetric
shell potential in a $D$-dimensional space. We have defined the potential shell at $R$ in terms of matching conditions imposed 
on the interior and exterior fields at $R$ that render the ensuing boundary value problem self-adjoint. In order to evaluate the Casimir energy of this configuration we have used the spectral zeta function regularization technique. Within this approach, the zeta function is expressed in terms of a contour integral of a characteristic function whose zeroes represent the eigenvalues of the problem. After a suitable analytic continuation 
we were able to compute the Casimir energy for scalar fields in a general shell potential configuration. In the previous section we analyzed explicitly the three-dimensional case by providing graphs of the energy as a function of specific parameters defining the matching conditions. 

Our results for the Casimir energy in the case $b=0$, which reduces to the well-known $\delta$-$\delta'$ potential, align, as expected, with those found previously in \cite{romaniega2022casimir}. The results presented for the case $b\neq 0$ are, on the other hand, new and reveal some interesting behavior of the Casimir energy in terms of the free parameters of the matching conditions. One of the most interesting aspects of the potentials with $b\neq 0$ is the observation that, unlike the case $b=0$, the Casimir energy can be negative for a specific range of values of the parameters that define the matching conditions. A negative Casimir energy translates into an inward pressure on the spherical shell.
Let us emphasize that by using the results for the Casimir energy in $D$ dimensions obtained in the previous section and a computer program such as Mathematica one can obtain specific results for any dimension $D$. There is an important caveat, however, that needs to be addressed.
For $D=3$ we were able to find, analytically, the range of values for the parameters $(a,b,c,d)$ which give a vanishing residue of the 
spectral zeta function at $s=-1/2$. In dimensions higher than $D=3$ however, such an analysis would be much more complicated and we are not able to predict whether analytic results, akin to the ones in the $D=3$ case, would even be possible. Despite this difficulty, a numerical analysis can always be performed to find the values of the parameters that give a vanishing residue and, hence, a well-defined energy.        

The work outlined in this paper can naturally be extended in a number of new directions. It would be interesting, for instance, to repeat the analysis presented here for the two-dimensional case. For $D=2$ the analytic continuation of the spectral zeta function would be slightly more involved for a couple of reasons. First, for $D=2$ and $l\geq 1$, the points $r=0$ and $r=\infty$ are limit points while for the zero angular mode $l=0$, the origin is, instead, a limit circle. This allows for generalized boundary conditions to be imposed at $r=0$. Second, the characteristic functions for $l=0$ would involve the modified Bessel function with index $0$ whose small argument expansion is different than the one for $K_{n}(z)$ and $I_{n}(z)$ with $n\geq 1$ \cite{watson,olver2010nist}. For this reason the contour deformation in the integral representation of the zeta function must be performed more carefully. The analysis of the spectral zeta function for a scalar field in two dimensions under the influence of a 
shell potential, would be quite interesting as the results could expand the set of exactly solvable models or be used in connection with conformal field theories.       

Another interesting extension of this work involves the asymptotic expansion of the trace of the heat kernel associated with the Laplacian endowed with a general shell-potential. In particular one can use the analytic continuation of the spectral zeta function obtained in this paper to compute explicitly the coefficients of the small-$t$ asymptotic expansion of the trace of the heat kernel in any dimension. The existence of the
small-$t$ asymptotic expansion of the trace of the heat kernel for the Laplacian endowed with shell potentials of the $\delta$-$\delta'$ type (namely $b=0$) has been established in \cite{1999heat}. As far as we know, a formal proof of the existence of the small-$t$ expansion of the trace of the heat kernel for the Laplacian with a general shell potential ($b\neq 0$) has not been explicitly provided and, hence, one cannot conclude that the residues of the associated spectral zeta function are indeed proportional to the coefficients of the heat kernel expansion.
In a future work, we hope to present a proof of the existence of the expansion of the trace of the heat kernel for general shell potentials and 
compute the coefficients in any dimension with the help of the spectral zeta function.


\appendix

\section{On the self-adjoint extensions of the radial operator}\label{app}

In this appendix we outline the two-interval self-adjoint extensions of the radial operator (\ref{radoper}). We follow the ideas 
described in \cite{zettl} where it is shown that the two-intervals self-adjoint extensions of a Sturm-Liouville operator are characterized by the 
boundary conditions imposed on the maximal domain functions. The types of boundary conditions allowed (if any) depend on whether an endpoint is  limit point, limit circle non-oscillatory or regular (for a definition of this classification see \cite{zettl} Chapter 7, Section 3 and \cite{everitt05}). 
We start the analysis by rewriting the eigenvalue equation associated with the radial operator (\ref{radoper}) in Sturm-Liouville form as
follows
\begin{equation}\label{app1a}
-\left(r^{D-1}\phi'\right)'+\left[\nu^{2}-\left(\frac{D-2}{2}\right)^{2}\right]r^{D-3}\phi=\lambda^{2}r^{D-1}\phi\;,
\end{equation}
where we have defined, for convenience $\nu=l+(D-2)/2$.  
The finite point $r=R$ is always a regular point, so we will focus on the points $r=0$ and $r=+\infty$. A numerically satisfactory pair of solutions of Bessel’s equation (\ref{app1a}) for $0<r<\infty$ is 
\begin{equation}
\phi_{1,\nu}(r)=r^{-\frac{D-2}{2}}J_{\nu}(r\sqrt{\lambda})\;,\quad\textrm{and}\quad \phi_{2,\nu}(r)=r^{-\frac{D-2}{2}}Y_{\nu}(r\sqrt{\lambda})\;.
\end{equation}
Since $\phi_{1,\nu}\notin L^{2}((a,\infty),r^{D-1}dr)$ with $a>0$, we can conclude that $r=+\infty$ is a limit point for
$\nu\geq 0$. For the point $r=0$, it is not difficult to realize that $\phi_{1,\nu}\in L^{2}((0,a),r^{D-1}dr)$ for $\nu\geq 0$ while 
$\phi_{2,\nu}\in L^{2}((0,a),r^{D-1}dr)$ for $\nu\in[0,1)$ and $\phi_{2,\nu}\notin L^{2}((0,a),r^{D-1}dr)$ when $\nu\geq 1$.
This implies that $r=0$ is a limit point when $\nu\geq 1$ (that is when $l\geq 1$ for $D\geq 2$ and when $l=0$ and $D\geq 4$) and is a limit circle non-oscillatory point 
when $\nu\in[0,1)$ (i.e. when $l=0$ and $D=2$ or $D=3$). 

According to the general Strurm-Liouville theory \cite{zettl}, ordinary boundary conditions can be imposed on functions in the maximal 
domain at any regular point. At a limit point, no boundary conditions are needed or allowed. Generalized boundary conditions need to be 
imposed, instead, at a limit circle non-oscillatory point. In order to impose generalized boundary conditions, one needs to introduce principal and non-principal solutions of (\ref{app1a}) (for a complete treatment of singular Sturm-Liouville problems and generalized boundary conditions 
see e.g. \cite{coddington} and \cite{dunford}). For $l=0$, a principal solution of (\ref{app1a}) is $f(r)=1$ when $D=2$ and $D=3$ while 
$\hat{f}(r)=\ln r$ is a non-principal solution when $D=2$ and $\hat{f}(r)=-1/r$ is a non-principal solution when $D=3$. 
Hence, when $r=0$ is in the limit circle non-oscillatory case, the generalized boundary conditions to be imposed on a function 
$\phi(r)\in D_{\textrm{max}}(h_{l})$ are
\begin{equation}\label{appbc1}
\tilde{\phi}(0)=\lim_{r\to 0}\frac{\phi(r)}{\ln r}\;,\quad \textrm{and} \quad \tilde{\phi}'(0)=\lim_{r\to 0}\left(\phi(r)-\tilde{\phi}(0)\ln r\right)\;,
\end{equation}          
for $D=2$, while when $D=3$ we have 
\begin{equation}\label{appbc2}
\tilde{\phi}(0)=-\lim_{r\to 0}r \phi(r)\;,\quad \textrm{and} \quad \tilde{\phi}'(0)=\lim_{r\to 0}\left(\phi(r)+\frac{\tilde{\phi}(0)}{r}\right)\;.
\end{equation}

We can now describe all the two-interval self-adjoint extensions of (\ref{radoper}) in terms of the boundary conditions that can be imposed on 
functions in $D_{\textrm{max}}(h_{l})$. It is convenient to distinguish two separate classes of two-intervals self-adjoint extensions depending on the nature of the origin: 
\begin{itemize}
\item \emph{$r=0$ is a limit point}. \newline
In this case the intervals $I_{<}$ and $I_{>}$ have two limit points, namely $r=0$ and $r=\infty$ where no boundary conditions are
allowed. This implies that all self-adjoint extensions are characterized by the ordinary boundary conditions imposed at the regular endpoint
$r=R$. In more detail, all self-adjoint extensions of (\ref{radoper}) in this two limit points class consist of functions $\phi(r)\in D_{\textrm{max}}(h_{l})$ that satisfy either \emph{separated} boundary conditions at $r=R$,
\begin{eqnarray}\label{bcsep}
&&\phi(R_{-})\cos\alpha+\phi'(R_{-})\sin\alpha=0\;,\nonumber\\
&&\phi(R_{+})\cos\beta+\phi'(R_{+})\sin\beta=0\;,
\end{eqnarray}
where $f(R_{\pm})=\lim_{r\to R^{\pm}}f(r)$, and $\alpha,\beta\in[0,\pi)$ or \emph{coupled} boundary conditions at $r=R$, i.e.
\begin{eqnarray}\label{bccoup}
\left(\begin{array}{c}
	 \phi(R_{+}) \\
	\phi'(R_{+})\\
	\end{array}\right)=e^{-i \mu}\Omega \left(\begin{array}{c}
	 \phi(R_{-}) \\
	\phi'(R_{-})\\
	\end{array}\right)\;,
\end{eqnarray}
with $\mu\in[0,\pi)$ and $\Omega\in SL(2,\mathbb{R})$ \cite{zettl}.

\item \emph{$r=0$ is a limit circle non-oscillatory point}. \newline
In this case, $r=\infty$ is the only limit point of the two intervals and the origin is a limit circle non-oscillatory point. This implies that generalized boundary conditions need to be imposed at $r=0$, while ordinary boundary conditions are imposed at $r=R$ (no boundary conditions are allowed at infinity). This implies that all two-intervals self-adjoint extensions of (\ref{radoper}) consist of functions $\phi(r)\in D_{\textrm{max}}(h_{l})$ that satisfy the following boundary conditions \cite{zettl}
\begin{equation}\label{bcthree}
{\cal A} \left(\begin{array}{c}
	 \tilde{\phi}(0) \\
	\tilde{\phi}'(0)\\
	\end{array}\right)+{\cal B}\left(\begin{array}{c}
	 \phi(R_{-}) \\
	\phi'(R_{-})\\
	\end{array}\right)+{\cal C}\left(\begin{array}{c}
	 \phi(R_{+}) \\
	\phi'(R_{+})\\
	\end{array}\right)=0\;,
\end{equation} 
where ${\cal A}$, ${\cal B}$ and ${\cal C}$ are $2\times 3$ matrices such that $\textrm{rank}({\cal A}|{\cal B}|{\cal C})=3$ and 
 \begin{eqnarray}\label{cond1}
&&\textrm{det}{\cal A}_{13}-\textrm{det}{\cal B}_{13}+\textrm{det}{\cal C}_{13}=0\;,\nonumber\\
&&\textrm{det}{\cal A}_{23}-\textrm{det}{\cal B}_{23}+\textrm{det}{\cal C}_{23}=0\;,\nonumber\\
&&\textrm{det}{\cal A}_{12}-\textrm{det}{\cal B}_{12}+\textrm{det}{\cal C}_{12}=0\;,
\end{eqnarray}
where we have assumed, for simplicity, that ${\cal A}$, ${\cal B}$ and ${\cal C}$ are real matrices and the notation ${\cal A}_{ij}$
denotes the $2\times 2$ submatrix of ${\cal A}$ constructed from the $i$-th and $j$-th rows of ${\cal A}$ (similarly for ${\cal B}_{ij}$
and ${\cal C}_{ij}$). 
\end{itemize} 

\section{Uniform asymptotic expansion of the characteristic function ${\cal F}_{\nu}(\nu z)$}\label{app1}

The characteristic function ${\cal F}_{\nu}(\nu z)$ is given in (\ref{charact1}) with coefficients defined in (\ref{coeffi}).
By utilizing the uniform asymptotic expansion of the modified Bessel functions and their derivative, which can be found in \cite{olver2010nist} 
Section 10.41, we find
\begin{equation}\label{asy1}
K_{\nu}(\nu z)I_{\nu}(\nu z)\sim\frac{t}{2\nu}\left(1+\sum_{k=1}^{\infty}\frac{C_{k}(t)}{\nu^{k}}\right)\;,
\end{equation}
\begin{equation}\label{asy0}
\left[K_{\nu}(\nu z)I_{\nu}(\nu z)\right]'\sim\frac{1}{\nu z}\sum_{k=1}^{\infty}\frac{B_{2k-1}(t)}{\nu^{2k-1}}\;,
\end{equation}
and
\begin{equation}\label{asy2}
K'_{\nu}(\nu z)I'_{\nu}(\nu z)\sim-\frac{1}{2\nu t z^{2}}\left(1+\sum_{k=1}^{\infty}\frac{\bar{C}_{k}(t)}{\nu^{k}}\right)\;,
\end{equation}
where we have defined $t=(1+z^{2})^{-1/2}$ and the polynomials in $t$ that appear in these expressions are
\begin{eqnarray}\label{CBC}
C_{k}(t)&=&(-1)^{k}\sum_{j=0}^{k}(-1)^{j}u_{j}(t)u_{k-j}(t)\;,\nonumber\\
B_{k}(t)&=&(-1)^{k}\sum_{j=0}^{k}(-1)^{j}v_{j}(t)u_{k-j}(t)\;,\nonumber\\
\bar{C}_{k}(t)&=&(-1)^{k}\sum_{j=0}^{k}(-1)^{j}v_{j}(t)v_{k-j}(t)\;.
\end{eqnarray}
The polynomials $u_{j}(t)$ and $v_{j}(t)$ are defined recursively, for $k\geq 0$, as (see \cite{olver2010nist}, Section 10.41)
\begin{eqnarray}
u_{k+1}(t)&=&\frac{1}{2}t^{2}(1-t^{2})u_{k}^{\prime}(t)+\frac{1}{8}\int_{0}^{t}(1-5\tau^{2})u_{k}(\tau)d\tau\;,\nonumber\\
v_{k+1}(t)&=&u_{k+1}(t)-\frac{1}{2}t(1-t^{2})\left[u_{k}(t)+2tu_{k}^{\prime}(t)\right]\;,
\end{eqnarray}
with $u_{0}(t)=v_{0}(t)=1$. It is interesting to notice that due to the symmetry of the expressions in (\ref{CBC}), it is not 
difficult to show that for $k\geq 0$
\begin{equation}\label{23}
C_{2k+1}(t)=\bar{C}_{2k+1}(t)=B_{2k}(t)=0\;.
\end{equation} 
We can now use the results (\ref{asy1})-(\ref{asy2}) to obtain the uniform asymptotic expansion
of the logarithm of the characteristic function ${\cal F}_{\nu}(\nu z)$. Due to the different leading behavior of the uniform asymptotic 
expansion, we will distinguish between two cases: $b\neq 0$ and $b=0$.

\paragraph{Case $b\neq 0$.}  In this case the uniform asymptotic expansion of ${\cal F}_{\nu}(\nu z)$ can be written as
\begin{eqnarray}\label{asybneq}
{\cal F}_{\nu}(\nu z)\sim \frac{b \nu}{2tR^{2}}\left(1+\sum_{n=1}^{\infty}\frac{{\cal D}_{n}(t)}{\nu^{n}}\right)\;,
\end{eqnarray}
where the polynomials ${\cal D}_{n}(t)$ can be found to be
\begin{eqnarray}
{\cal D}_{1}(t)&=&\frac{2tR^{2}\delta}{b}\;,\nonumber\\
{\cal D}_{2}(t)&=&\frac{\alpha t^{2}R^{2}}{b}+\frac{2\gamma tR}{b}B_{1}(t)+\bar{C}_{2}(t)\;,\nonumber\\
{\cal D}_{n}(t)&=&\bar{C}_{n}(t)+\frac{\alpha t^{2}R^{2}}{b}C_{n-2}(t)+\frac{2\gamma tR}{b}B_{n-1}(t)\;,\quad n\geq3\;.
\end{eqnarray}
The relation (\ref{23}) allows us to conclude that for $n\geq 1$, we have ${\cal D}_{2n+1}(t)=0$. 
By using (\ref{asybneq}) one obtains an expansion for the logarithm of the characteristic function, i.e.
\begin{equation}\label{asymcharb}
\ln{\cal F}_{\nu}(\nu z)\sim \ln\left(\frac{b\nu}{2R^{2}}\right)-\ln t+\sum_{n=	1}^{\infty}\frac{\Omega_{n}(t)}{\nu^{n}}\;,
\end{equation}
where the polynomials $\Omega_{n}(t)$ are defined through the cumulant expansion
\begin{equation}
\ln\left(1+\sum_{n=1}^{\infty}\frac{{\cal D}_{n}(t)}{\nu^{n}}\right)\sim \sum_{n=1}^{\infty}\frac{\Omega_{n}(t)}{\nu^{n}}\;,
\end{equation}
and have the form
\begin{equation}\label{om}
\Omega_{n}(t)=\sum_{l=0}^{2n}\omega_{n,l+n}t^{l+n}\;.
\end{equation}
The explicit expressions of the polynomials $\Omega_{n}(t)$ for the first few values of $n$ are
\begin{equation}
\Omega_{1}(t)=\frac{2 \delta  R^2 t}{b}\;,
\end{equation}
\begin{equation}
\Omega_{2}(t)=t^2 \left[\frac{R \left(-b \gamma +\alpha  b R-2 \delta ^2 R^3\right)}{b^2}-\frac{3}{8}\right]+t^4 \left(\frac{\gamma  R}{b}+\frac{5}{4}\right)-\frac{7 t^6}{8}\;,
\end{equation}
\begin{equation}
\Omega_{3}(t)=\frac{t^3 \left[3 b \delta  R^2 (3 b+8 R (\gamma -\alpha  R))+32 \delta ^3 R^6\right]}{12 b^3}-\frac{t^5 \left[\delta  R^2 (5 b+4 \gamma  R)\right]}{2 b^2}+\frac{7 \delta  R^2 t^7}{4 b}\;,
\end{equation}
\begin{eqnarray}
\Omega_{4}(t)&=&t^6 \left[\frac{4 \delta ^2 R^4 (5 b+4 \gamma  R)+b R (b (23 \gamma -8 \alpha  R)+4 \gamma  R (\gamma -\alpha  R))}{4 b^3}+\frac{109}{16}\right]\nonumber\\
&+&\frac{t^8 \left[-733 b^2+8 b R (6 \alpha  R-41 \gamma )-16 R^2 \left(\gamma ^2+7 \delta ^2 R^2\right)\right]}{32 b^2}\nonumber\\
&-&\frac{t^4 \left[b^2 \left(27 b^2+16 b R (3 \gamma -2 \alpha  R)+32 R^2 (\gamma -\alpha  R)^2\right)+32 b \delta ^2 R^4 (3 b+8 R (\gamma -\alpha  R))+256 \delta ^4 R^8\right]}{64 b^4}\nonumber\\
&+&t^{10} \left(\frac{21 \gamma  R}{4 b}+\frac{441}{16}\right)-\frac{707 t^{12}}{64}\;.
\end{eqnarray}
From these expressions one is immediately able to read off the coefficients $\omega_{n,l+n}$ in (\ref{om}).
\paragraph{Case $b=0$.} Since the matrix representing the coupled boundary conditions in (\ref{bccoupexp}) must satisfy the 
condition $ad-bc=1$, when $b=0$ one must necessarily have $\delta\neq 0$. This remark together with the asymptotic expansions (\ref{asy1}) and
(\ref{asy0}) lead to 
\begin{eqnarray}\label{asybzero}
{\cal F}_{\nu}(\nu z)\sim \delta\left(1+\sum_{n=1}^{\infty}\frac{\tilde{{\cal D}}_{n}(t)}{\nu^{n}}\right)\;,
\end{eqnarray}
where the polynomials $\tilde{{\cal D}}_{n}(t)$ are
\begin{eqnarray}
\tilde{{\cal D}}_{1}(t)&=&\frac{\tilde{\alpha}t}{2}+\frac{\tilde{\gamma}}{R}B_{1}(t)\;,\nonumber\\
\tilde{{\cal D}}_{n}(t)&=&\frac{\tilde{\alpha}t}{2}C_{n-1}(t)+\frac{\tilde{\gamma}}{R}B_{n}(t)\;,\quad n\geq2\;,
\end{eqnarray}
with the coefficients $\tilde{\alpha}=\alpha/\delta$ and $\tilde{\gamma}=\gamma/\delta$. Similarly to the previous case, one can prove, by 
using (\ref{23}), that for $n\geq 1$, $\tilde{{\cal D}}_{2n}(t)=0$. The corresponding expansion for the logarithm of the characteristic function
is
\begin{equation}\label{asymchar0}
\ln{\cal F}_{\nu}(\nu z)\sim \ln\delta+\sum_{n=1}^{\infty}\frac{\tilde{\Omega}_{n}(t)}{\nu^{n}}\;,
\end{equation}
where the polynomials $\tilde{\Omega}_{n}(t)$ are defined through the cumulant expansion
\begin{equation}
\ln\left(1+\sum_{n=1}^{\infty}\frac{\tilde{{\cal D}}_{n}(t)}{\nu^{n}}\right)\sim \sum_{n=1}^{\infty}\frac{\tilde{\Omega}_{n}(t)}{\nu^{n}}\;,
\end{equation}
and have the form
\begin{equation}\label{om1}
\tilde{\Omega}_{n}(t)=\sum_{l=0}^{2n}\tilde{\omega}_{n,l+n}t^{l+n}\;.
\end{equation} 
The explicit expressions of the polynomials $\tilde{\Omega}_{n}(t)$ for the first few values of $n$ are
\begin{equation}
\tilde{\Omega}_{1}(t)=\frac{t^3 \tilde{\gamma }}{2 R}+\frac{t \left(R \tilde{\alpha }-\tilde{\gamma }\right)}{2 R}\;,
\end{equation}
\begin{equation}
\tilde{\Omega}_{2}(t)=-\frac{t^6 \tilde{\gamma }^2}{8 R^2}-\frac{t^4 \left(2 R \tilde{\alpha } \tilde{\gamma }-2 \tilde{\gamma }^2\right)}{8 R^2}-\frac{t^2 \left(\tilde{\gamma }^2+R^2 \tilde{\alpha }^2-2 R \tilde{\alpha } \tilde{\gamma }\right)}{8 R^2}\;,
\end{equation}
\begin{eqnarray}
\tilde{\Omega}_{3}(t)&=&\frac{t^9 \left(2 \tilde{\gamma }^3+105 R^2 \tilde{\gamma }\right)}{48 R^3}+\frac{t^7 \left(-2 \tilde{\gamma }^3+5 R^3 \tilde{\alpha }-65 R^2 \tilde{\gamma }+2 R \tilde{\alpha } \tilde{\gamma }^2\right)}{16 R^3}\nonumber\\
&+&\frac{t^5 \left[2 \tilde{\gamma }^3-6 R^3 \tilde{\alpha }+R^2 \left(2 \tilde{\alpha }^2+33\right) \tilde{\gamma }-4 R \tilde{\alpha } \tilde{\gamma }^2\right]}{16 R^3}\nonumber\\
&+&\frac{t^3 \left[-2 \tilde{\gamma }^3+R^3 \tilde{\alpha } \left(2 \tilde{\alpha }^2+3\right)-3 R^2 \left(2 \tilde{\alpha }^2+3\right) \tilde{\gamma }+6 R \tilde{\alpha } \tilde{\gamma }^2\right]}{48 R^3}\;,
\end{eqnarray}
\begin{eqnarray}
\tilde{\Omega}_{4}(t)&=&-\frac{t^{12} \left(\tilde{\gamma }^4+70 R^2 \tilde{\gamma }^2\right)}{64 R^4}+\frac{t^{10} \tilde{\gamma } \left[\tilde{\gamma }^3-R \tilde{\alpha } \left(\tilde{\gamma }^2+20 R^2\right)+50 R^2 \tilde{\gamma }\right]}{16 R^4}\nonumber\\
&-&\frac{t^8 \left[3 \tilde{\gamma }^4+5 R^4 \tilde{\alpha }^2-76 R^3 \tilde{\alpha } \tilde{\gamma }+R^2 \left(3 \tilde{\alpha }^2+98\right) \tilde{\gamma }^2-6 R \tilde{\alpha } \tilde{\gamma }^3\right]}{32 R^4}\nonumber\\
&+&\frac{t^6 \left[\tilde{\gamma }^4+3 R^4 \tilde{\alpha }^2-R^3 \tilde{\alpha } \left(\tilde{\alpha }^2+20\right) \tilde{\gamma }+3 R^2 \left(\tilde{\alpha }^2+6\right) \tilde{\gamma }^2-3 R \tilde{\alpha } \tilde{\gamma }^3\right]}{16 R^4}\nonumber\\
&-&\frac{t^4 \left(R \tilde{\alpha }-\tilde{\gamma }\right) \left[-\tilde{\gamma }^3+R^3 \tilde{\alpha } \left(\tilde{\alpha }^2+2\right)-3 R^2 \left(\tilde{\alpha }^2+2\right) \tilde{\gamma }+3 R \tilde{\alpha } \tilde{\gamma }^2\right]}{64 R^4}\;.
\end{eqnarray}
Just like in the case $b\neq 0$, these expressions can be used to extract the coefficients $\tilde{\omega}_{n,l+n}$ in (\ref{om1}).

\section{An identity for some generalized Bernoulli polynomials}\label{app2}

The aim of this Appendix is to prove the relation $B_{n-1}^{(n)}(j)=0$ for $j=1,\ldots,n-1$ which has been used in Section \ref{casEner} 
during the computation of the residue of the spectral zeta function at $s=-1/2$. Remarkably, we have not been able to find a proof of the aforementioned relation after a search of the relevant literature on generalized Bernoulli polynomials.

The generalized Bernoulli polynomials $B_{k}^{\alpha}(x)$ are defined in terms of the relation \cite{srivastava88}
\begin{equation}\label{app21}
\left(\frac{z}{e^{z}-1}\right)^{\alpha}e^{x z}=\sum_{j=0}^{\infty}B_{j}^{(\alpha)}(x)\frac{z^j}{j!}\;,\quad |z|<2\pi\;.
\end{equation}
By setting $\alpha=n\in\mathbb{N}^{+}$ and $x=l$ with $l=1,\ldots,n-1$, and by using the following series expansion defining the N\"{o}rlund 
polynomials \cite{carlitz60} for $|z|<2\pi$,
\begin{equation}
\left(\frac{z}{e^{z}-1}\right)^{n}=\sum_{m=0}^{\infty}B_{m}^{(n)}\frac{z^m}{m!}\;,
\end{equation}
the expression in (\ref{app21}) can be rewritten as
\begin{equation}
e^{lz}\sum_{m=0}^{\infty}B_{m}^{(n)}\frac{z^m}{m!}=\sum_{j=0}^{\infty}B_{j}^{(n)}(l)\frac{z^j}{j!}\;.
\end{equation}

By expressing $e^{lz}$ in terms of its Taylor expansion and by then using the Cauchy product to rewrite the product of the two resulting 
power series in terms of a single one, we obtain, for the term proportional to the $z^{n-1}$ power, the following expression
\begin{equation}\label{app22}
B_{n-1}^{(n)}(l)=\sum_{p=0}^{n-1}\binom{n-1}{p}B_{p}^{(n)}l^{n-1-p}\;.
\end{equation}   
Since the following relation holds between the N\"{o}rlund polynomials and the Stirling numbers \cite{carlitz60}
\begin{equation}
\binom{n-1}{p}B_{p}^{(n)}=s(n,n-p)\;,
\end{equation} 
we can re-express (\ref{app22}) as
\begin{equation}
B_{n-1}^{(n)}(l)=\sum_{j=0}^{n-1}s(n,j+1)l^j\;.
\end{equation}

By shifting the summation index in the last expression and by noticing that $s(n,0)=0$ for $n\in\mathbb{N}^{+}$, we finally obtain,
for $l=1,\ldots,n-1$,
\begin{equation}
B_{n-1}^{(n)}(l)=\frac{1}{l}\sum_{j=0}^{n}s(n,j)l^j=0\;,
\end{equation}
since \cite{olver2010nist} (formula 26.8.7)
\begin{equation}
\sum_{k=0}^{n}s\left(n,k\right)x^{k}=x(x-1)(x-2)\cdots(x-n+1)\;.
\end{equation}

\bibliography{BibliographyV1}{}
\bibliographystyle{unsrt}

\end{document}